\documentclass[times, twocolumn]{aastex63}
\usepackage{CJK}
\usepackage{graphicx}
\usepackage{enumerate}
\usepackage{amssymb, amsmath}
\usepackage{natbib}
\usepackage{color}
\usepackage{ulem}
\usepackage{dblfnote}
\usepackage{appendix}
\usepackage{hyperref}
\usepackage[stable]{footmisc}
\usepackage{comment}
\usepackage{footnote}
\usepackage[nodayofweek]{datetime}
\newdateformat{monthyearday}{%
  \THEYEAR\ \monthname[\THEMONTH] \twodigit{\THEDAY}}

\submitjournal{AAS Journals}
\shorttitle{Keck/OSIRIS-Pa$\beta$ PDS~70}
\shortauthors{Uyama et al.}

\begin{document}

\title{Keck/OSIRIS Pa$\beta$ high-contrast imaging and updated constraints on PDS~70b}

%% Note that the corresponding author command and emails has to come
%% before everything else. Also place all the emails in the \email
%% command instead of using multiple \email calls.
\correspondingauthor{Taichi Uyama}
\email{tuyama@ipac.caltech.edu}

\author[0000-0002-6879-3030]{Taichi Uyama}
    \affiliation{Infrared Processing and Analysis Center, California Institute of Technology, 1200 E. California Blvd., Pasadena, CA 91125, USA}
    \affiliation{NASA Exoplanet Science Institute, Pasadena, CA 91125, USA}
    \affiliation{National Astronomical Observatory of Japan, 2-21-1 Osawa, Mitaka, Tokyo 181-8588, Japan}

\author[0000-0002-6318-0104]{Chen Xie}
    \affiliation{Aix Marseille Univ, CNRS, CNES, LAM, Marseille, France}
    
\author[0000-0003-0568-9225]{Yuhiko Aoyama}
    \affiliation{Institute for Advanced Study, Tsinghua University, Beijing 100084, People's Republic of China}
    \affiliation{Department of Astronomy, Tsinghua University, Beijing 100084, People's Republic of China}
    
% main contributors - alphabetical

\author[0000-0002-5627-5471]{Charles A. Beichman}
    \affiliation{NASA Exoplanet Science Institute, Pasadena, CA 91125, USA}
    \affiliation{Infrared Processing and Analysis Center, California Institute of Technology, 1200 E. California Blvd., Pasadena, CA 91125, USA}

\author[0000-0002-3053-3575]{Jun Hashimoto}
    \affiliation{Astrobiology Center, 2-21-1 Osawa, Mitaka, Tokyo 181-8588, Japan}
    \affiliation{Subaru Telescope, National Astronomical Observatory of Japan, Mitaka, Tokyo 181-8588, Japan}

% other collaborators - alphabetical

\author[0000-0001-9290-7846]{Ruobing Dong}
    \affiliation{Department of Physics \& Astronomy, University of Victoria, Victoria, BC, V8P 1A1, Canada}

\author[0000-0002-9017-3663]{Yasuhiro Hasegawa}
    \affiliation{Jet Propulsion Laboratory, California Institute of Technology, 4800 Oak Grove Dr., Pasadena, CA, 91109, USA}

\author[0000-0002-5658-5971]{Masahiro Ikoma}
    \affiliation{National Astronomical Observatory of Japan, 2-21-1 Osawa, Mitaka, Tokyo 181-8588, Japan}

\author[0000-0002-8895-4735]{Dimitri Mawet}
    \affiliation{Department of Astronomy, California Institute of Technology, 1200 E. California Blvd., Pasadena, CA 91125, USA}
    \affiliation{Jet Propulsion Laboratory, California Institute of Technology, 4800 Oak Grove Dr., Pasadena, CA, 91109, USA}

\author[0000-0003-0241-8956]{Michael W. McElwain}
    \affiliation{Exoplanets and Stellar Astrophysics Laboratory, Code 667, Goddard Space Flight Center, Greenbelt, MD 20771, USA}

\author[0000-0003-2233-4821]{Jean-Baptiste Ruffio}
\affiliation{Department of Astronomy, California Institute of Technology, 1200 E. California Blvd., Pasadena, CA 91125, USA}

\author[0000-0002-4309-6343]{Kevin R. Wagner}
    \altaffiliation{NASA Hubble Fellowship Program -- Sagan Fellow}
    \affiliation{Steward Observatory, University of Arizona, Tucson, AZ 85721, USA}
    \affiliation{NASA NExSS Alien Earths Team, USA}

\author[0000-0003-0774-6502]{Jason J. Wang}
    \altaffiliation{51 Pegasi b Fellow}
    \affiliation{Department of Astronomy, California Institute of Technology, 1200 E. California Blvd., Pasadena, CA 91125, USA}

\author[0000-0003-2969-6040]{Yifan Zhou}
    \altaffiliation{Harlan J. Smith McDonald Observatory Fellow}
    \affiliation{Department of Astronomy/McDonald Observatory, The University of Texas, 2515 Speedway, Stop C1400 Austin, TX 78712, USA}

\begin{abstract}
We present a high-contrast imaging search for Pa$\beta$ line emission from protoplanets in the PDS~70 system with Keck/OSIRIS integral field spectroscopy. We applied the high-resolution spectral differential imaging technique to the OSIRIS $J$-band data but did not detect the Pa$\beta$ line at the level predicted using the parameters of \cite{Hashimoto2020}. This lack of Pa$\beta$ emission suggests the MUSE-based study may have overestimated the line width of H$\alpha$. We compared our Pa$\beta$ detection limits with the previous H$\alpha$ flux and H$\beta$ limits and estimated $A_{\rm V}$ to be $\sim0.9$ and 2.0 for PDS~70~b and c respectively. In particular, PDS~70~b's $A_{\rm V}$ is much smaller than implied by high-contrast near-infrared studies, which suggests the infrared-continuum photosphere and the hydrogen-emitting regions exist at different heights above the forming planet.
\end{abstract}


\section{Introduction} \label{sec: Introduction}

A variety of theoretical and observational studies have investigated planet formation, yet the mechanisms are still poorly understood. 
High-contrast imaging at infrared (IR) wavelengths can detect the thermal emission of young exoplanets directly and thus provide key insights to distinguish between various planet formation mechanisms. Characterization of the physical and atmospheric parameters of protoplanets at specific ages helps in assessing the initial conditions of their formation \citep[e.g.][]{Bonnefoy2014}. Furthermore, addressing such problems as reconciling the evolutionary cooling models \citep[hot/warm/cold-start;][]{Spiegel2012} with the relevant physical processes \citep[e.g. core accretion and disk instability;][]{Pollack1996,Boss1997} are essential to improving our understanding of planet formation.  One of the ways to probing planet formation is to observe hydrogen emission originating in active mass accretion onto protoplanets \citep[][]{Aoyama2018}.

PDS~70 is one of the most intriguing young systems with high-contrast imaging, revealing two protoplanets located within a large cavity of the protoplanetary disk \citep[PDS~70bc;][]{Keppler2018,Haffert2019} and follow-up observations confirming active mass accretion onto them \citep[e.g.][]{Haffert2019}.
Previous studies have explored some of the hydrogen emission lines in the PDS~70 system; H$\alpha$ (656.28~nm), H$\beta$ (486.14~nm), Br$\alpha$ (4.050~\micron), and Br$\gamma$ (2.166~\micron). 
H$\alpha$ emission has been reported by MagAO \citep{Wagner2018}, VLT/MUSE \citep{Haffert2019}, and HST \citep{Zhou2021}. The measured H$\alpha$ flux shows temporal variability on a 1--2-year timescale for reasons which are still controversial: either systematic instrumental calibration errors and/or an intrinsic time variability. 
The MUSE data include H$\beta$ line but yielded only a null detection \citep{Hashimoto2020} with 3$\sigma$ upper limits of 2.3 and 1.6$\times10^{-16}$ erg~s$^{-1}$~cm$^{-2}$ for PDS~70~b and c respectively.
\cite{Christiaens2019} reported the $K$-band spectrum of PDS~70~b taken by VLT/SINFONI ({\it R}$\sim$100) and \cite{Wang2021} presented the $K$-band spectra of PDS~70~bc taken by VLT/GRAVITY (MEDIUM resolution), but they did not detect significant Br$\gamma$ emission.
\cite{Wang2021} set 3$\sigma$ upper limits of Br$\gamma$ to 5.1 and 4.0$\times10^{-17}$ erg~s$^{-1}$~cm$^{-2}$
%4.0$\times10^{-20}$ W~m$^{-2}$ {\bf [YH: it would be better to use cgs, which is used above]}
for PDS~70~b and c respectively, which are limited by the $K$-band continua of PDS~70~bc.
\cite{Stolker2020} reported the detection of PDS~70~bc with VLT/NACO NB4.05 filter (Br$\alpha$ filter; $\lambda_{\rm cen}$=4.05\micron, $\Delta\lambda$=0.02\micron). However, they suggested that PDS~70~b's spectrum is best fit by an atmospheric model without Br$\alpha$ and did not argue in favor of a line detection. In addition to the hydrogen emission lines, \cite{Zhou2021} reported ultraviolet (UV) emission from PDS~70~b with HST/WFC F336W filter and suggested that the hydrogen continuum emission dominates the UV flux.
\cite{Aoyama2020} incorporated all these lines into a discussion of the emission mechanisms but were unable to  determine fully the physical and accretion parameters of PDS~70~bc.

Here we report on a search for the previously unobserved line of Pa$\beta$ (1.282~\micron) around PDS~70 which is one of the brightest emission lines relative to H$\alpha$. We used Keck/OSIRIS mid-resolution integral field spectroscopy (IFS, {\it R}$\sim$4000) to further investigate the accretion mechanisms of PDS~70~bc.
The observations and preliminary result of the post-processing were originally reported in \cite{Uyama2021}. In this paper we present the updated results with a detailed analysis of the data following \cite{Xie2020} (see Sections~\ref{sec: Data} and \ref{sec: Results}). Section~\ref{sec: Discussions} investigates constraints on the accreting parameters of PDS~70~bc by incorporating the OSIRIS results with the previous studies.

\section{Data} \label{sec: Data}
\subsection{Observations} \label{sec: Observations}

We observed PDS~70 with Keck/OSIRIS in the Jn3-band on 2020 May 31 UT (PI: Charles Beichman) to search for a Pa$\beta$ emission line (1.282~ $\micron$) from accretion onto the protoplanets. We used the OSIRIS IFS spatial sampling of 0\farcs02 spaxel$^{-1}$ that covers a field of view of 0\farcs96~$\times$~1\farcs28, where each spatial location has a spectrum from 1.275 $\micron$ to 1.339 $\micron$ (Jn3) with resolving power of $\sim$4000. The observations achieved a total exposure time of 4800~sec (120-sec single exposure $\times$ 40 frames) under good seeing conditions (0\farcs4--0\farcs6). The typical full width half maximum (FWHM) of the PDS~70's point spread function (PSF) measured a diffraction-limited $\sim$60--70 mas, but the quality of the last sequence of the observations was poor because of high airmass ($>$2.2) and relatively bad seeing ($\sim0\farcs7$). 
Hence we excluded the last eight frames from this analysis. 
By taking the ratio of the flux within a 3-by-3-spaxel aperture and within the entire field-of-view (FoV), we estimated the Strehl ratio to be 9.88\% at Pa$\beta$. Due to the relatively small FoV (0\farcs96~$\times$~1\farcs28), we may overestimate the Strehl ratio. The low Strehl ratio (typically $<20\%$) can lead to flux loss and we took into account this effect in the data analysis.
We also obtained unsaturated images of HD~143956 \cite[spectral type: B9;][]{Houk1988} and HD~144609 \cite[spectral type: K0;][]{Houk1988} for telluric correction and photometric reference, respectively. The details of the OSIRIS observations can be found in Table~\ref{tab:obs_log}.

% %%%%%%%%%%%%%%%%%%%%%%%%%%%%%%%%%%%%%%%%%%%%%%%%
\begin{deluxetable}{ccc}
\label{tab:obs_log}      % is used to refer this table in the text
% \tablenum{1}
\tablecaption{OSIRIS observations using the Jn3 filter with the plate scale of 20 mas.}
\tablewidth{0pt}
\tablehead{
\colhead{Target} & \colhead{$t_{\rm DIT} \times n_{\rm DIT}$$^{a}$} & \colhead{On-source time}\\
\colhead{ } & \colhead{ } & \colhead{(s)}
}
\startdata
PDS~70 &  40$\times$ 120 &  4800$^{(b)}$  \\
HD~143956 & 20 $\times$ 1 &  20   \\
HD~144609 & 2 $\times$ 1 &  2   \\
\enddata
\tablecomments{$^{a}$~$t_{\rm DIT}$ is the exposure time per image frame in the 
unit of seconds and $n_{\rm DIT}$ is the number of image frames.$^{b}$ The last eight frames were excluded in the analysis due to the inferior observing conditions, resulting in a practical total integration time of 3840~s.}
\end{deluxetable}

\subsection{Data Reduction} \label{sec: Data Reduction}
We used the OSIRIS Data Reduction Pipeline \citep[reduction type: astronomical reduction pipeline;][]{Lyke2017, Lockhart2019} with the corresponding rectification matrices\footnote{\url{http://tkserver.keck.hawaii.edu/osiris/}} to extract the data cube and calibrated for dark subtraction, cosmic-ray removal, telluric correction, and wavelength solution.    To search for faint companions with single emission lines, we need to first subtract the stellar light accurately.  The preliminary data reduction presented in \cite{Uyama2021} applied the PCA-based SDI reduction that was originally used for the MUSE data \citep{Hashimoto2020}. However, this reduction technique left some instrumental residuals due to sensitivity differences between the OSIRIS spaxels.
We therefore applied an advanced high-resolution spectral differential imaging (HRSDI) technique to remove the stellar emission (see, \cite{Haffert2019} and \cite{Xie2020} for the details). HRSDI is suitable for retrieving sharp emission lines while removing the stellar halo. 
However, before we applied the HRSDI to the final combined dataset, some residual bad pixels were removed from each exposure that passed through the OSIRIS Data Reduction Pipeline. To remove the bad pixels, we first applied HRSDI on each exposure, aiming for reducing the influence of stellar emission in the next step. Next, we applied a sigma clipping algorithm on the dithered exposures to make a bad pixel mask for each exposure. Then all the exposures were centered on the flux peak and mean combined after the removal of bad pixels. 

The process of HRSDI consists of two steps, removing the stellar emission and removing the uncalibrated instrumental effects. The stellar emission was subtracted from all normalized spaxels with the normalized reference spectrum that was obtained after the continuum-normalization \citep{Haffert2019}. 
The uncalibrated instrumental residuals were removed using a principal component analysis (PCA) subtraction technique \citep{Soummer2012, Amara_Quanz2012}. For example, the instrumental residual in \cite{Uyama2021} can be removed with the first few PCA components. The number of PCA components to subtract was determined by maximizing the signal-to-noise ratio (S/N) of injected fake planets at the location of PDS~70~b (see also, Section \ref{sec: Fake Planet Injection}). 

\subsection{Fake Planet Injection} \label{sec: Fake Planet Injection}

To estimate the instrumental throughput, we performed the fake planet injection described in \cite{Xie2020}. The instrumental throughput includes the flux loss due to the low Strehl ratio (see Section \ref{sec: Observations}) and that made by the PSF subtraction. Unless we specifically mentioned, both effects were corrected throughout the paper.
The fake planet was created based on a planet spectrum and a stellar PSF. We used a single Gaussian line as the planet spectrum because our observations did not utilize angular differential imaging and thus did not achieve sufficient contrast to detect the continua of PDS~70~bc.
We adopted the line-of-sight redshift of 25~km~s$^{-1}$ \citep{Haffert2019} and a FWHM of 70~km~s$^{-1}$ or 0.3~nm. 
The injected Gaussian line can be covered by two spectral channels. We measured the flux using the aperture photometry in spectral channels of 1281.75~nm and 1281.90~nm with a square aperture of 3 by 3 spaxels (60$\times$60~mas). 
The noise was estimated at the same spatial location in the spectral direction after HRSDI, using 150 spectral channels (bandwidth: 22.5 nm) around Pa$\beta$. After obtaining a 5~$\sigma$ detection, we estimated the flux loss caused by the PSF subtraction by comparing the injected and recovered flux. 
The flux losses caused by the PSF subtraction are 28\% and 14\% at the location of PDS~70~b and c, respectively.

\section{Results} \label{sec: Results}
\begin{figure}
    \centering
    \includegraphics[width=0.45\textwidth]{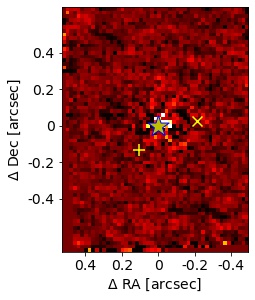}
    \caption{The HRSDI-reduced Keck/OSIRIS data at wavelengths of 1281.75 and 1281.9~nm (combined image). The locations of PDS~70, PDS~70~b, and c are indicated by the star, plus, and cross symbols respectively.}
    \label{fig: SDI image}
\end{figure}

\begin{figure*}
\begin{tabular}{cc}
\begin{minipage}{0.5\hsize}
    \centering
    \includegraphics[width=0.9\textwidth]{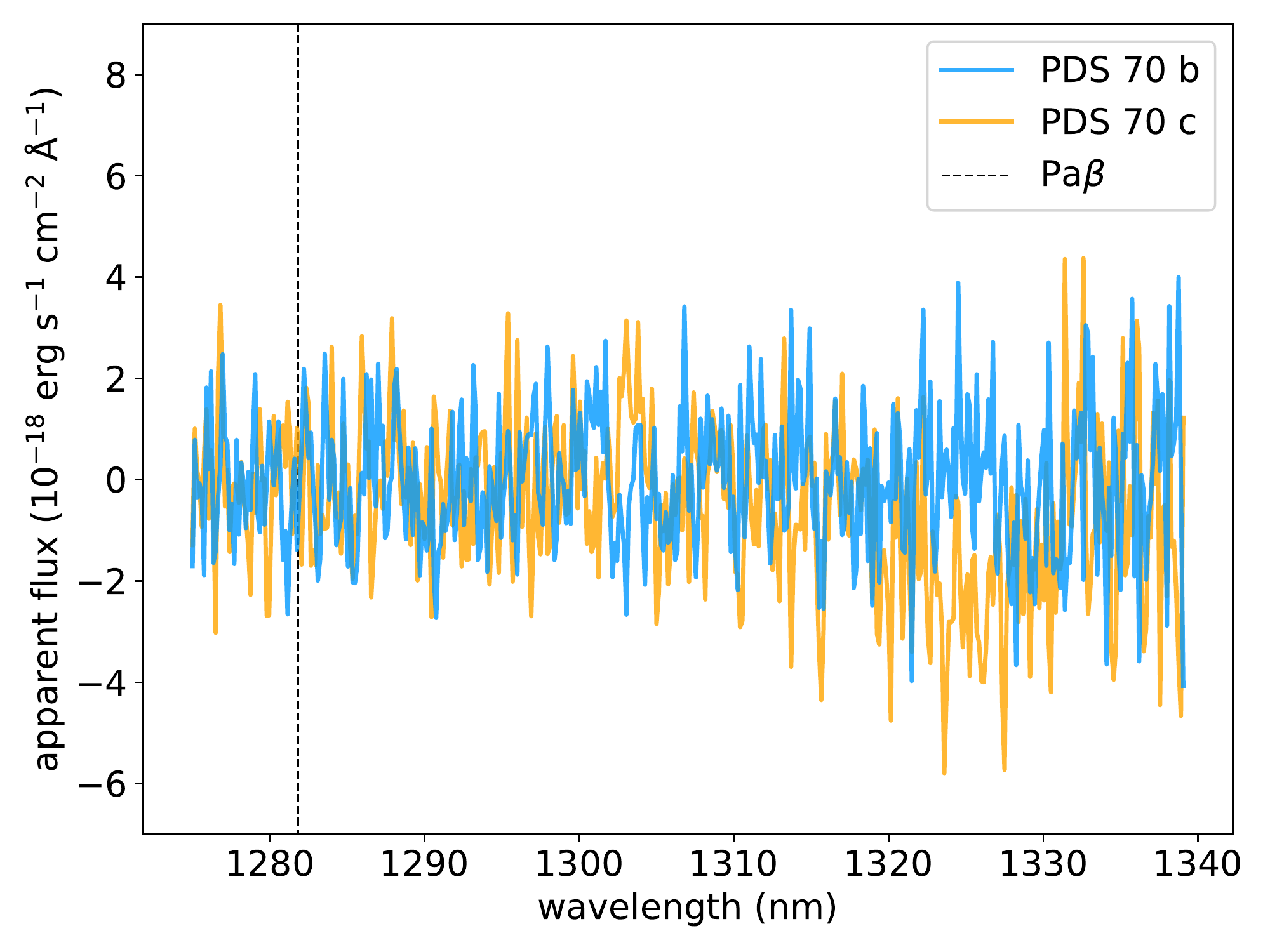}
    \caption{Spectra of the residuals after the HRSDI reduction at the locations of PDS~70~b and c. For display purposes, no throughput correction was made.} 
    \label{fig: SDI spectra}
\end{minipage}
\begin{minipage}{0.5\hsize}
    \centering
    \includegraphics[width=0.9\textwidth]{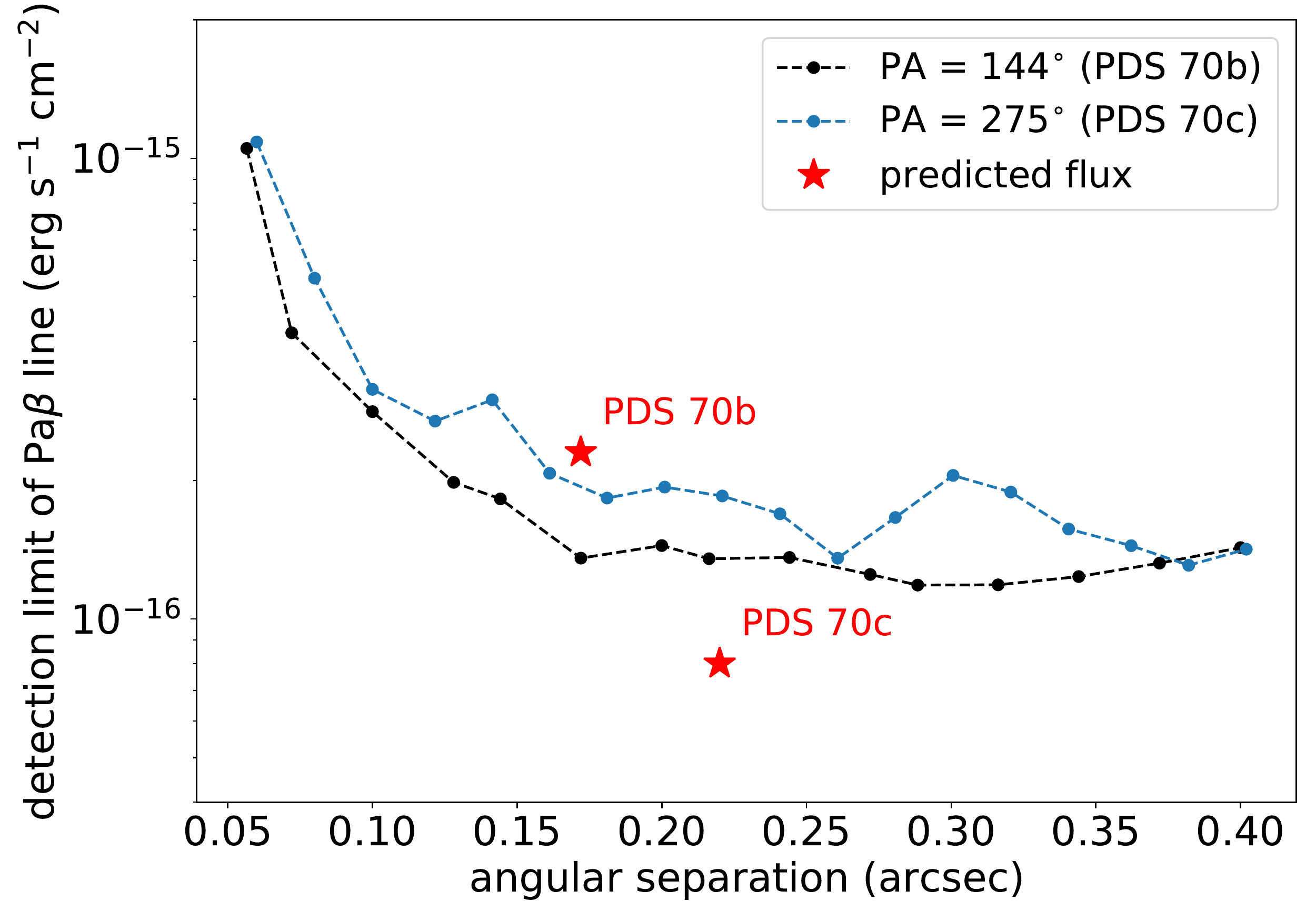}
    \caption{Radial profiles of 5$\sigma$ detection limits at two position angles. The predicted Pa$\beta$ fluxes of PDS~70~bc assuming the estimated parameters of \cite{Hashimoto2020} are indicated by the red stars. }
    \label{fig: Pab limit}
\end{minipage}
\end{tabular}
\end{figure*}

After the post-processing as mentioned in Section~\ref{sec: Data Reduction} we did not detect Pa$\beta$ at the locations of PDS~70~b and c (see Figure \ref{fig: SDI image}).
Figure \ref{fig: SDI spectra} shows the residual spectra after the HRSDI reduction at the location of PDS~70bc. We then calculated the 5$\sigma$ detection limits\footnote{The 5$\sigma$ detection limit is defined as the summation of the flux in the aperture on the residual image and 5 times of the corresponding noise. As mentioned in Section~\ref{sec: Fake Planet Injection}, the estimated noise (without throughput correction) at the locations of PDS~70~b and c are $2.5 \times 10^{-18}$ erg~s$^{-1}$~cm$^{-2}$ and $2.7 \times 10^{-18}$ erg~s$^{-1}$~cm$^{-2}$, respectively. The residual fluxes at the locations of PDS~70~b and c are $-2.7 \times 10^{-18}$ erg~s$^{-1}$~cm$^{-2}$ and $2.5 \times 10^{-18}$ erg~s$^{-1}$~cm$^{-2}$, respectively.} of $1.4 \times 10^{-16}$ erg~s$^{-1}$~cm$^{-2}$ and 1.9 $\times 10^{-16}$ erg~s$^{-1}$~cm$^{-2}$ for PDS~70b and c, respectively. The correction of the flux loss caused by the PSF subtraction and the low Strehl ratio has been taken into account. Figure~\ref{fig: Pab limit} shows the radial profiles for 5$\sigma$ detection limits at the two position angles (PAs) of the two planets. We note that the PSF of OSIRIS is not circularly symmetric. Although PDS~70~c is further away from the star, the noise at the location of PDS~70~c is higher, resulting in a higher detection limit.

\cite{Uyama2021} defined the noise as a standard deviation of a spectral channel at the location of PDS~70~b after the SDI reduction without taking into account the OSIRIS' spectral resolution and flux loss by the post-processing. Their calculations also used the literature value of PDS~70 $J$-band flux \citep[$J$=9.553 mag;][]{2MASS} to convert the contrast limit into a flux detection limit, but the central star is variable due to its activity and potentially also veiling by the circumstellar disk. In this study we used a field star of HD~144609 \citep[$J$=5.459 mag;][]{2MASS} as a photometric reference and calculated a conversion factor from ADU to the apparent flux.

We also investigated the validity of the estimated limits by injecting fake sources. We used %\cite{Aoyama2018} 
\cite{Aoyama2019} to convert the MUSE-based H$\alpha$ profiles into the Pa$\beta$ profiles assuming the derived parameters of PDS~70~bc (the number density: $n_0=3.8\times10^{12}$ cm$^{-3}$, the gas velocity: $v_0=144$ km~s$^{-1}$, and the extinction: $A_{\rm H_\alpha}=2.4$ mag) in \cite{Hashimoto2020}. 
Our prediction for the Pa$\beta$ flux from PDS~70~b is comparable to the actual OSIRIS detection limit. Since we did not detect Pa$\beta$ emission our model may have overestimated the Pa$\beta$ flux. Alternatively, \cite{Hashimoto2020} may have overestimated the 10\% and 50\% widths of the H$\alpha$ profiles and thus the parameters of $n_0$ and/or $v_0$, possibly because MUSE does not have sufficient spectral resolution ({\it R}$\sim$2500).
This latter interpretation can explain the difference between the mass measurements from the IR SED \citep[e.g.][]{Wang2020,Stolker2020} and the hydrogen emission lines \citep[][]{Hashimoto2020}. The mass estimate in \cite{Hashimoto2020} using the \cite{Aoyama2019} model is an upper limit on the dynamical mass of PDS~70~b.

\section{Discussion} \label{sec: Discussions}

We use our detection limits of Pa$\beta$ to further constrain the physical parameters of PDS~70~bc with a theoretical model \citep{Aoyama2018,Aoyama2019}. 
For a comparison with our Pa$\beta$ detection limits, we refer to: 1) the MUSE results \citep{Hashimoto2020} which is most similar to the OSIRIS data rather than MagAO or HST because of the similarity of the data format and post-processing techniques; and 2) the HST results \citep{Zhou2021} which were obtained in 2020 May close to when we observed PDS~70 with OSIRIS, thereby mitigating any effects of the year-timescale intrinsic variability.

We assume magnetospheric accretion (filling factor of the hydrogen emission - the coverage fraction of the shock on the planetary surface: $f_{\rm f}\lesssim 0.1$) for the accretion mechanism of PDS~70~bc \citep[e.g.][]{Thanthibodee2019}, from which we can set a lower limit in the ($n_0$, $v_0$) parameter space \citep[see also Figure 3 in][for the modeled H$\alpha$ luminosity with different filling factor values]{Hashimoto2020}. With this assumption, H$\beta$ and Pa$\beta$ line strengths are expected to be close to the MUSE and OSIRIS detection limits, respectively. Detailed explanations about the relationship between filling factor, other accretion parameters, and hydrogen emission luminosity are given in \cite{Aoyama2020}. 
If the filling factor is much larger than the above assumption we cannot simply compare the Pa$\beta$ limits with the theoretical model. For example, when the shock comes from the circumplanetary disk surface flow rather than the magnetospheric accretion, the filling factor is a few tens of percent \citep{Takasao+2021}.

\subsection{Comparison between the OSIRIS and MUSE results} \label{sec: Comparison between the OSIRIS and MUSE results}
Instrumental differences in the comparison of the visible and IR data should be small since the MUSE and OSIRIS IFSs have similar properties and the two datasets were treated in a similar fashion, using HRSDI to remove the stellar halo and searching for emission lines at small angular separations. We compare our Pa$\beta$ detection limits with the MUSE-based H$\alpha$ fluxes. However, we note that we have the uncertainty of time variability due to the difference of the epochs. 

\begin{figure*}
\begin{tabular}{cc}
\begin{minipage}{0.5\hsize}
    \centering
    \includegraphics[width=0.9\textwidth]{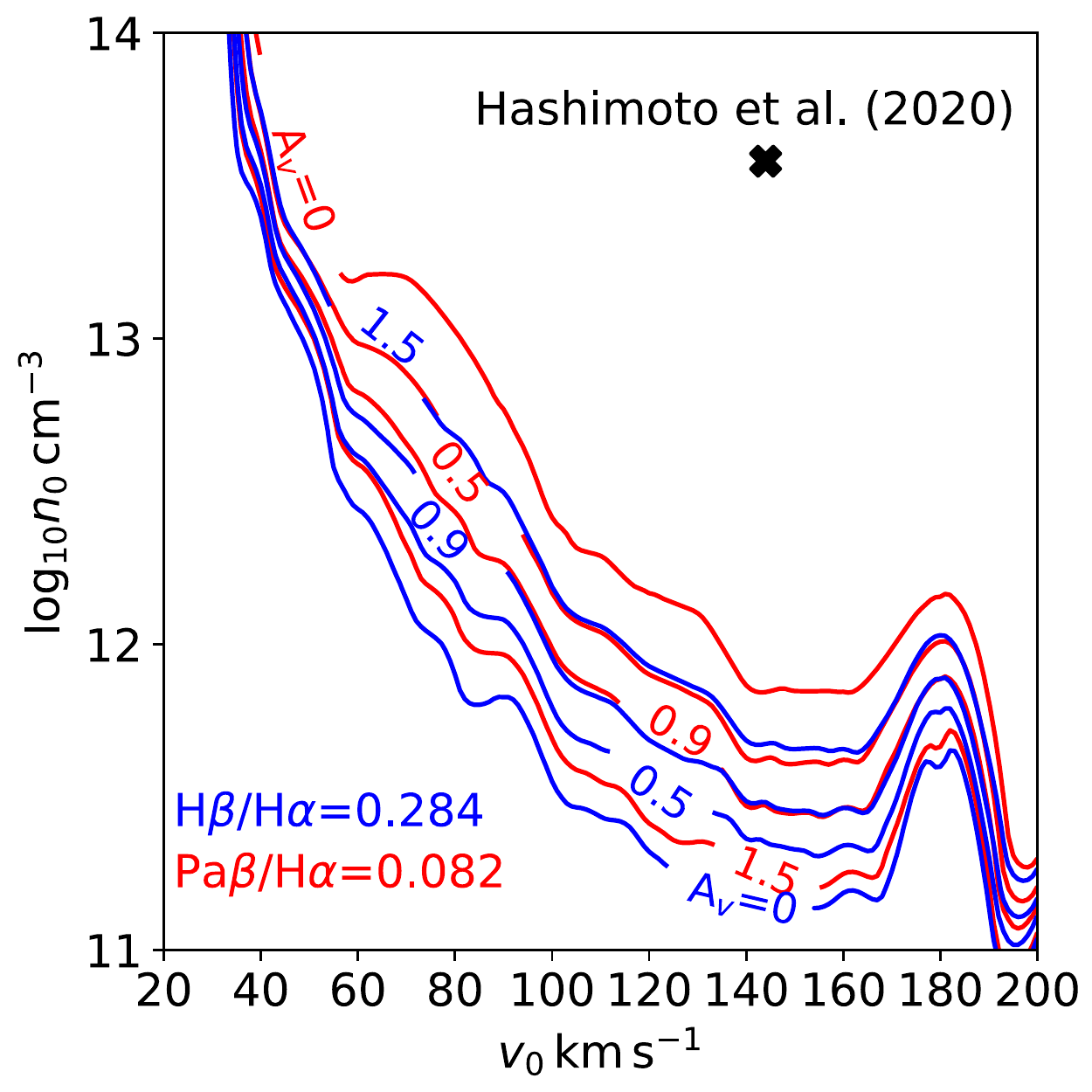}
\end{minipage}
\begin{minipage}{0.5\hsize}
    \centering
    \includegraphics[width=0.9\textwidth]{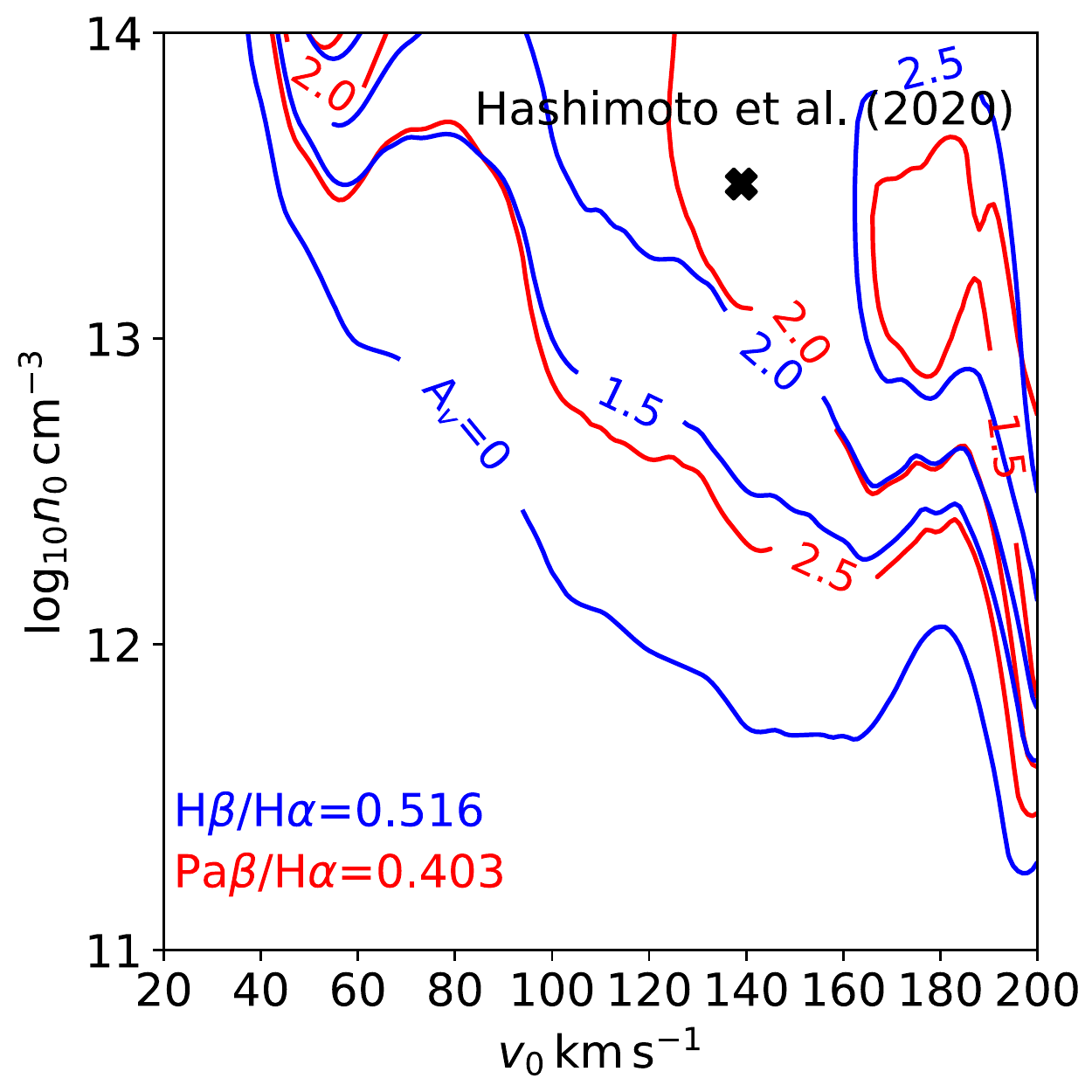}
\end{minipage}
\end{tabular}
 \caption{Contours of the 3$\sigma$ H$\beta$ detection limits \citep[blue;][]{Hashimoto2020} and the 3$\sigma$ Pa$\beta$ detection limits (red, this work) in comparison with the MUSE-based H$\alpha$ flux of PDS~70~b (left) and PDS~70~c (right). The estimated $n_0$ and $v_0$ of PDS~70bc in \cite{Hashimoto2020} are indicated as black crosses respectively.
 We take into account the extinction effect ($A_{\rm V}$) and the wavelength dependency \citep[see Equations (9) and (10) in][]{Wang2019extinction}. As H$\beta$ is bluer and Pa$\beta$ is redder than H$\alpha$, using these detection limits enables us to set upper and lower limits for $A_{\rm V}$, from which we estimate $A_{\rm V}$. }
    \label{fig: contours MUSE}
\end{figure*}

We used our 3$\sigma$ Pa$\beta$ detection limits ($6.6\times10^{-17}$ erg~s$^{-1}$~cm$^{-2}$ and $1.3\times10^{-16}$ erg~s$^{-1}$~cm$^{-2}$ for PDS~70~b and c respectively) and the MUSE-based H$\alpha$ fluxes and 3$\sigma$ H$\beta$ limits \citep{Hashimoto2020} to constrain the PDS~70~bc's parameters. Combining the hydrogen line data from these two AO-fed integral field units provides better constraints on the effects of extinction.
The difference of extinction effect between H$\beta$/H$\alpha$ and Pa$\beta$/H$\alpha$ ratios enables us to estimate the $A_{\rm V}$ value. Figure~\ref{fig: contours MUSE} shows the contours of line flux ratio as a function of $n_0$ and $v_0$, with a variety of $A_{\rm V}$ values for PDS~70~b (left) and c (right) respectively. Although our final detection limit is higher than the preliminary result presented in \cite{Uyama2021}, the comparison between the Pa$\beta$ and H$\beta$ limits suggests that $A_{\rm V}$ for \textit{the line emitting region of PDS~70~b} is consistent with $\sim0.9$ \citep[$A_{\rm H\alpha}\sim0.69$~mag assuming the extinction law in][]{Wang2019extinction}. 

Our extinction estimates are lower than other estimates. \cite{Hashimoto2020} attributed the failure of MUSE to detect H$\beta$ to large extinction ($A_{\rm H\alpha}>2.0$~mag) but this may be due to the overestimation of ($n_0$, $v_0$) and due to the insufficient spectral resolution of MUSE as mentioned in Section \ref{sec: Results}. 
Our derived $A_{\rm V}$ value is also inconsistent with the Spectral Energy Distribution (SED)-fitting argument with the GRAVITY observations \citep[$A_{\rm V}\sim4-10$~mag assuming ISM extinction and the best-fit extincted models;][]{Wang2021} that used the shape of the continuum and the molecular-mapping argument from SINFONI observations \citep[$A_{\rm V}\sim16-17$~mag ;][]{Cugno2021}, which used the depths of the lines.
However, this discrepancy might suggest a vertical difference between the location of the  photosphere responsible for the IR continuum and the hydrogen-emitting regions. The evaporated materials at the shock can sublimate beneath the hydrogen-emitting regions to create an additional extinction source for the PDS~70~b's atmosphere. 
This assumption does not conflict with the physical assumption of \cite{Aoyama2018}. In that sense, IR-continuum observations and hydrogen-emission observations of protoplanets should be careful to identify each extinction effect independently.
The large difference between ($n_0$, $v_0$) estimated in \cite{Hashimoto2020} and the ($n_0$, $v_0$) contour with $A_{\rm V}=0.9$~mag suggests that the filling factor may be larger than a lower limit of \cite{Hashimoto2020} ($f_{\rm f}\gtrsim0.01$) by about an order of magnitude. To test the hypothesis about the filling factor, observing the hydrogen emission line with higher spectral resolution is required.
We note that the discussion in this section ignores the time variability as mentioned above. Section \ref{sec: Comparison between the OSIRIS and HST results} takes into account the variability effect.

For the case of PDS~70~c, we could not explore as deep parameter space as PDS~70~b because the Pa$\beta$ detection limit is higher than that of PDS~70~b as mentioned in Section \ref{sec: Results} and the H$\alpha$ flux is smaller \citep{Haffert2019,Hashimoto2020}.
The comparison with the Pa$\beta$ and H$\beta$ detection limits suggests $A_{\rm V}\sim2.0$~mag ($A_{\rm H\alpha}\sim1.5$~mag). Compared with \cite{Hashimoto2020}, who set a lower limit of $A_{\rm H\alpha}$ and $f_{\rm f}$ to 1.1~mag and $\sim0.003$ respectively, our estimated $A_{\rm V}$ value is consistent with their argument.
We note that this argument relies on the assumption that the H$\alpha$ profile of PDS~70~c was sufficiently resolved by MUSE. 
If $n_0$ and $v_0$ of PDS~70~c given in \cite{Hashimoto2020} are overestimated as well as those of PDS~70~b, higher contrast levels at H$\beta$ and Pa$\beta$ are required to constrain these parameters.
As mentioned above, PDS~70~c's photospheric continuum may also be extincted by additional material compared with the hydrogen-emitting regions.
Resolving the H$\alpha$ line profile with higher resolution and/or deeper searches for H$\beta$ and Pa$\beta$ will improve the constraints on the accretion parameters of PDS~70~c.

\subsection{Comparison between the OSIRIS and HST results} \label{sec: Comparison between the OSIRIS and HST results}

\begin{figure}
    \centering
    \includegraphics[width=0.45\textwidth]{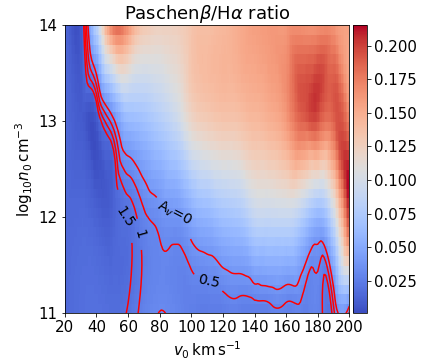}
    \caption{3$\sigma$ Pa$\beta$ detection limit (red) in comparison with the HST-based H$\alpha$ flux of PDS~70~b. The contours assume $A_{\rm V}$= 0, 0.5, 1.0, and 1.5 (from top to bottom) and the same wavelength dependency as Figure \ref{fig: contours MUSE}.}
    \label{fig: contours HST}
\end{figure}

As mentioned above, the OSIRIS and MUSE observations were not conducted in the same epoch and thus simply comparing these observational results leaves the uncertainty of the temporal variability. 
\cite{Zhou2021} monitored PDS~70~b's H$\alpha$ line with HST between 2020 February and 2020 July, which covers the OSIRIS observation on 2020 May 31 UT, and did not find larger variability in the line flux than 30\% ($\sim2.4\sigma$). They also suggested the hydrogen line emission was variable on a year-timescale by incorporating  MagAO and MUSE results obtained in 2018 \citep{Wagner2018,Hashimoto2020}.
In this section we compare our Pa$\beta$ detection limit of PDS~70~b with the time-averaged H$\alpha$ flux estimated from the HST observations \citep[$1.62\pm0.23\times10^{-15}\ {\rm erg\ s^{-1}\ cm^{-2}}$;][]{Zhou2021}. Although the HST data format and post-processing technique are different from OSIRIS, we used injection testing to account for differences in instrumental parameters and data analysis techniques.
Figure \ref{fig: contours HST} shows the same contours of PDS~70~b as Figure \ref{fig: contours MUSE} assuming our Pa$\beta$ limit and the HST-based H$\alpha$ flux. Note that we do not include the H$\beta$ limits because the H$\beta$ observations were not conducted at the same epoch. 
The higher H$\alpha$ flux value than the MUSE result helps us to explore a deeper parameter space. 
Assuming that the extinction effect is stable at $A_{\rm V}=0.9$~mag, our 3$\sigma$ detection limit can set an upper limit of $v_0$ at $\sim70$~km~s$^{-1}$. Using Equation (3) in \cite{Hashimoto2020}, this upper limit corresponds to $\sim$3-4~$M_{\rm Jup}$ for the upper limit of PDS~70~b's mass and is consistent with the mass estimation by the IR high-contrast studies \citep[e.g.]{Wang2020,Stolker2020}.
To better determine/constrain the (variable) accreting parameters simultaneous observations of H$\alpha$, H$\beta$, and Pa$\beta$ are more helpful.
%The contours do not change significantly compared with Figure \ref{fig: contours MUSE}, which may suggest the year-timescale variation does not affect at least the $A_{\rm V}$ estimation.

\citet{Zhou2021} estimated the continuum flux at the wavelength $\lambda = 336$\,nm to be $(1.4\pm0.3)\times10^{-18}\,\mathrm{erg\,s^{-1}\,cm^{-2}\,\AA^{-1}}$. This wavelength is located in the hydrogen Balmer continuum.
Using the model of \citet{Aoyama2018}, we can estimate the fluxes of the hydrogen recombination continua from the shock-heated gas, as a byproduct of the hydrogen line fluxes.
The model prediction can reproduce both the continuum and H$\alpha$ fluxes observed for PDS~70~b, with some parameter sets. However, our calculation with ($v_0,\,n_0$)=($144\,\mathrm{km\,s^{-1}}, 3.8\times10^{12}\,\mathrm{cm^{-3}}$), which is estimated in \citet{Hashimoto2020}, resulted in $F_{\lambda,336}/F_\mathrm{H\alpha}=5.2\times10^{-3}\,\mathrm{\AA^{-1}}$, where $F_{\lambda,336}$ is the flux per unit wavelength at $\lambda=336$\,nm and $F_\mathrm{H\alpha}$ is the H$\alpha$ flux, while its observed value is $(8.6\pm2.2)\times10^{-4}\,\mathrm{\AA^{-1}}$ when the flux in the F656N filter of HST represents the H$\alpha$ flux \citep{Zhou2021}.
This comparison %without extinction taken into account indicates that the emission from PDS~70~b is subject to no strong extinction, which is 
shows inconsistency with the results of \citet{Hashimoto2020}. This implies that the spectral profile given by MUSE can be overestimated, which is consistent with our interpretation about the null detection of Pa$\beta$ in the OSIRIS observations. Note that the above estimate of $(v_0, n_0)$ comes from the MUSE-based spectral profile.
However, the continuum flux for higher values of $v_0$ is less reliable due to a lack of coolants effective for hot gases in the \cite{Aoyama2018} model. Also, a part of photosphere that is heated by the accretion should emit continuum \citep[e.g.,][]{Hartmann2016}.
Further theoretical studies on planetary recombination continua are essential.

\section{Summary} \label{sec: Summary}
We present high-contrast spectral imaging for the unexplored emission line of Pa$\beta$ from PDS~70~bc with Keck/OSIRIS integral field spectroscopy. After removing stellar halo utilizing the same HRSDI technique that was applied to VLT/MUSE observations, we did not detect Pa$\beta$ despite the predicted Pa$\beta$ flux of PDS~70~b from the estimated accretion parameters in \cite{Hashimoto2020} being comparable to the detection limit of our dataset. The null detection suggests that our model overestimated the Pa$\beta$ flux, probably because MUSE does not have sufficient spectral resolution and \cite{Hashimoto2020} overestimated $n_0$ and $v_0$ from the H$\alpha$ profile.

We then compared our detection limits with previous H$\alpha$ and H$\beta$ observations to set further constraints on the accretion parameters. We adopted two H$\alpha$ observations from MUSE and HST - comparing OSIRIS with MUSE can assume the smallest systematic difference in terms of the data format and post-processing techniques, whereas HST covers 2020 May when we observed PDS~70 thereby minimizing the effect of time variability on our conclusions.
The MUSE-based comparison between Pa$\beta$/H$\alpha$ and H$\beta$/H$\alpha$ ratios enables us to estimate $A_{\rm V}$ assuming the extinction law. We estimated $A_{\rm V}\sim0.9\ {\rm and}\ 2.0$ for PDS~70~bc respectively. Particularly the derived $A_{\rm V}$ of PDS~70~b is inconsistent with the previous NIR studies, but this might suggest an additional extinction source of PDS~70~b's IR-continuum photosphere that is located beneath the hydrogen emitting regions.
The HST-based Pa$\beta$/H$\alpha$ ratio suggested that the year-timescale variation does not significantly affect the $A_{\rm V}$ estimate. We also incorporated the Balmar continuum detected by HST/WFC F336W observations in the \cite{Aoyama2018} framework. The comparison between the Balmer continuum with H$\alpha$ flux suggests that the H$\alpha$ spectral profile may be overestimated. This interpretation is consistent with the null detection of Pa$\beta$ in our OSIRIS observations.

Higher spectral resolution will resolve the hydrogen emission line profiles and a deeper search could detect multiple hydrogen emissions, which helps to better estimate the accreting parameters and understand the accretion mechanisms of protoplanets. 

\newpage
\acknowledgments
We thank Jim Lyke for kindly helping to confirm the OSIRIS FoV settings.
The authors would like to thank the anonymous referees for their constructive comments and suggestions to improve the quality of the paper.
T.U. is supported by Grant-in-Aid for Japan Society for the Promotion of Science (JSPS) Fellows and JSPS KAKENHI Grant No. JP21J01220.
Y.H. is supported by the Jet Propulsion Laboratory, California Institute of Technology, under a contract with NASA.
K.W. acknowledges support from NASA through the NASA Hubble Fellowship grant HST-HF2-51472.001-A awarded by the Space Telescope Science Institute, which is operated by the Association of Universities for Research in Astronomy, Incorporated, under NASA contract NAS5-26555.
The results reported herein benefited from collaborations and/or information exchange within NASA's Nexus for Exoplanet System Science (NExSS) research coordination network sponsored by NASA's Science Mission Directorate.

The data presented herein were obtained at the W. M. Keck Observatory, which is operated as a scientific partnership among the California Institute of Technology, the University of California and the National Aeronautics and Space Administration. The Observatory was made possible by the generous financial support of the W. M. Keck Foundation.
This publication makes use of data products from the Two Micron All Sky Survey, which is a joint project of the University of Massachusetts and the Infrared Processing and Analysis Center/California Institute of Technology, funded by the National Aeronautics and Space Administration and the National Science Foundation.

We wish to acknowledge the critical importance of the current and recent Mauna Kea Observatories daycrew, technicians, telescope operators, computer support, and office staff employees, especially during the challenging times presented by the COVID-19 pandemic. Their expertise, ingenuity, and dedication is indispensable to the continued successful operation of these observatories.
The authors wish to recognize and acknowledge the very significant cultural role and reverence that the summit of Maunakea has always had within the indigenous Hawaiian community.  We are most fortunate to have the opportunity to conduct observations from this mountain.

\newpage
\bibliography{library}                                    
%\end{CJK*}
\end{document}